# IMAGE DENOISING BY MEDIAN FILTER IN WAVELET DOMAIN

Afrah Ramadhan[1], Firas Mahmood[2] and Atilla Elci[3]

[1]Dept. of Electrical - Electronics Engineering, Aksaray University, Aksaray, Turkey
[2] Technical College of Engineering, Duhok Polytechnic University, Duhok, Iraq
[3]Dept. of Electrical - Electronics Engineering, Aksaray University, Aksaray, Turkey

## ABSTRACT

*The details of an image with noise may be restored by removing noise through a suitable image de-noising method. In this research, a new method of image de-noising based on using median filter (MF) in the wavelet domain is proposed and tested. Various types of wavelet transform filters are used in conjunction with median filter in experimenting with the proposed approach in order to obtain better results for image de-noising process, and, consequently to select the best suited filter. Wavelet transform working on the frequencies of sub-bands split from an image is a powerful method for analysis of images. According to this experimental work, the proposed method presents better results than using only wavelet transform or median filter alone. The MSE and PSNR values are used for measuring the improvement in de-noised images.*

## 1. INTRODUCTION

The need for efficient image de-noising techniques has grown with the huge production of digital images. No matter how good cameras are, image improvement is often desirable to extend their range of action [2]. A digital image is often corrupted by some noise native in the production process and equipment, or injected into the image during its transmission. There are different types of noise appearing in images, such as impulse noise, uniform noise, Salt-and-Pepper and additive Gaussian noise [2, 3, 6]. Noise is present in an image either in an additive or multiplicative form [11] as formulated below:

$$\text{Additive noise} \quad : w(x, y) = s(x, y) + n(x, y)$$
$$\text{Multiplicative noise} \quad : w(x, y) = s(x, y) \times n(x, y)$$

where $s(x,y)$ is the original image, $n(x,y)$ is the noise introduced into the image producing the corrupted image $w(x, y)$, and $(x,y)$ represents the pixel location [10,11,12]. Figure 1 shows the de-noising concept by applied linear operation noise $n(x,y)$ added to the image $s(x,y)$ to form the noisy image $w(x,y)$, then applied de-noising technique to produce the restored image $z(x,y)$.

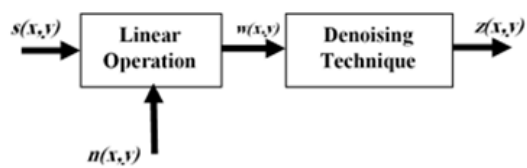

Figure 1: Noising and de-noising concept.

The best way to test the effect of noise on a standard digital image is to insert Additive Gaussian Noise AGN [2] as the possibility of seeing this type of noise in real life images is greater than other types [2, 3]. AGN is used in this research; and, noisy image with AGN becomes more difficult during image de-noising process.

The Gaussian noise is defined by the following probability density function F of Gaussian random variable $g$ in Eq. 1 [10, 11, 12].

$$F(g) = \frac{1}{\sqrt{2\pi\sigma^2}} \, e^{\frac{-(g-m)^2}{2\sigma^2}} \quad \dots 1 \, [8]$$

where:

g  : gray level
m  : mean or average of the function
σ2  : variance of the noise

The proposed de-noising scheme is being introduced in the following section; its four variations also considered in search of excellence. Section 3 takes up experimental testing of the proposed schemes. Section 4 provides discussion and Section 5 conclusions.

## 2. PROPOSED DE-NOISING SCHEME

In digital image processing, de-noising technique used in order to reduce, or if possible remove, noise without losing detail. There are various de-noising techniques. Gaussian noise can be reduced using a spatial filter. While smoothing an image, the blurring of fine-scaled image edges and details is possible for they correspond to the high frequencies blocked by the process. Among the spatial filtering techniques conventionally employed for noise removal are mean (convolution) filtering, median filtering and Gaussian smoothing. Most of the standard algorithms used for de-noising perform an individual filtering process. The result decreases the noise, but the image is blurred or smoothed at lines and edges due to high frequency losses.

In order to effectively reduce the noise with minimum adverse outcome, this research examines employing various filters and wavelet transform in analyzing a digital image in useful sub-bands. Wavelet transform and filtering, as well as a median filter are widely used individually for image de-noising. In this work, we explore combinations that would produce better outcome for digital image de-noising.

Digital image de-noising using Discrete Wavelet Transform (DWT) approach is being highlighted in the following steps:

1. Apply DWT on the noisy image to split it into four sub-bands (A, H, V, D) by using wavelet filter families [6, 4].
2. Choose one of the suitable threshold rule methods to compute a threshold value for each one of the detail bands H, V, D.
3. Compare pixels in the sub-bands (H, V, D) against the specific threshold value for that band. Set the pixel to zero if its value is less than the threshold of the band; otherwise, test the next pixel value. This step gets repeated for all the pixels of the selected sub-band.
4. Apply inverse DWT (IDWT) on the sub-bands to obtain the de-noisy (de-noised) image.
5. Apply median filter within WT alone before threshold or after threshold as needed by in each case. Median filter follow the moving window principle. A 3×3 filter mask of pixels is scanned over pixel matrix of the entire image, sorting all the values of the pixel from the surrounding neighborhood in an ascending manner. Exchange the pixel with the middle pixel value [3].

6. Compare the de-noised and the original image by computing the image quality of the de-noising technique. The result expressed as the values of the Mean Square Error (MSE) and the Peak Signal to Noise Ratio (PSNR). MSE and PSNR values are computed using the equations 2 and 3 respectively [13, 14].

$$MSE = \frac{1}{M.N} \sum_{i=0}^{M-1} \sum_{j=0}^{N-1} [F(i,j) - I(i,j)]^2 \qquad \text{.................... 2}$$

$$PSNR = 10. \log_{10} \frac{(max)^2}{MSE} \qquad \text{........................3}$$

where:
I (i, j)   : original image
F (i, j)   : de-noised image
M and N: size of the original image
max     : maximum pixel value of grayscale image that is used in this work which equals to 255.

Thresholding is a simple and effective method of image segmentation used to generate digital images from a grayscale image. As such, threshold technique is used beneficially for digital image de-noising. Threshold technique is used in this study in digital wavelet domain. As it is also dependent on the capabilities of wavelet transform for removing noise, the best choice of threshold value would provide great de-noising [3, 7]. The selection of the threshold value is an important factor for it determines which pixel values are to be kept or deleted in a sub-band. In this study one level of DWT decomposition is applied and the threshold value is compared against all the pixels at each detail bands (H, V, and D). Therefore, a good threshold value would lead to less distortion of the image [3, 5, 7].

Threshold value is computed by applying the equation 4 as follows:

$$Th = \frac{\sigma^2}{\sigma_x} \qquad \text{........... 4}$$

Where:

$\sigma^2$: the noise value of noisy image, which is computed by using Robust Median Estimator estimated from the sub-band HH by equation 5 below.

$$\sigma^2 = [median(|x(i,j)|)/0.6745]^2 \qquad \text{............…….. 5}$$

$\sigma_x$ : is the standard deviation (STD) of the sub-bands, computed in the equation 6 [8].

$$\sigma = \sqrt{\frac{\sum_{i=0}^{n-1}(xi-m)^2}{n}} \qquad \text{………….. 6}$$

Where:
$\sigma$    : The standard deviation (STD)
m      : the mean
n      : the number of pixels in the sub-bands
xi      : sub-band.

Threshold plays a major role in removal of noise from images because de-noising most frequently produces smoothed images, also reducing their sharpness [7, 9, 15]. Table 1 displays the estimated values of noise ratio, obtained by applying equation 5 at various noise levels.

Table 1: Estimated values of the noise ratio for sample images using HH sub-band (results of applying Robust Median Estimator)

| Image Name | Std original image | Std Noise=15 | Std Noise=20 | Std Noise=25 | Std Estimation Noise=15 | Std Estimation Noise=20 | Std Estimation Noise=25 |
|---|---|---|---|---|---|---|---|
| Barbara | 54.9897 | 57.3279 | 58.7963 | 60.2864 | 18.3562 | 25.0196 | 28.0826 |
| Camera | 63.3729 | 64.3469 | 65.0112 | 65.9484 | 17.6443 | 22.8960 | 27.4976 |
| Butterfly | 48.0703 | 50.5352 | 52.3114 | 54.3618 | 15.2861 | 23.0569 | 28.8463 |

In this paper, two algorithms are considered for digital image de-noising based on using median filter combined with DWT: the two classic de-noising methods depending on applying the DWT only or the median filter only and mixing between the two techniques. In the following, the flow diagrams illustrate each test case of the image de-noising methods. According to the analysis of these sub-bands, digital image de-noising algorithm is applicable on the following four cases.

## 2.1. THE FIRST CASE: APPLYING DWT ONLY

In this case, only DWT is applied as de-noising technique in order to obtain sub-bands (Approximation band, A, and detail bands H, V, and D). Equation 5 is used to estimate the noise in the sub-band D, equation 6 is used to determine standard deviation for the detail sub-bands, then equation 4 is used to find the threshold value for each sub-band. Pixel values of each sub-band are compared against the band threshold value. If the pixel value is greater than the threshold value, then it is set to zero. Otherwise, next pixel is processed (see equation 7). This process is repeated until all the pixels are handled. At this time, IDWT is applied on de-noised sub-bands. MSE and PSNR values are computed using equations 2 and 3 on the de-noised and the original images. A block diagram of the process flow for this case is being displayed in Figure 2.

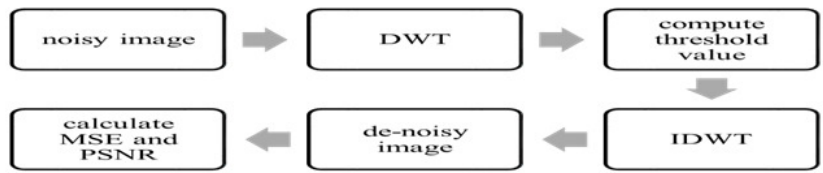

Figure 2: The block diagram for the first case, DWT only

## 2.2. THE SECOND CASE: APPLYING MEDIAN FILTER ONLY

In this case, only median filter is used to apply de-noising technique on the noisy image, consequently obtaining the de-noised image then calculate MSE and PSNR as in case one. Figure 3 shows the block diagram for this second proposed method.

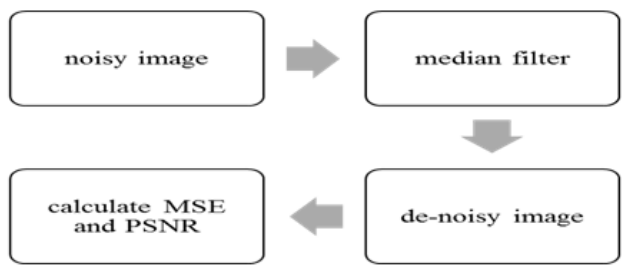

Figure 3: The block diagram for the second case, Median filter only

## 2.3. THE THIRD CASE: APPLYING MEDIAN FILTER BEFORE COMPUTING THE THRESHOLD VALUE

In this case, DWT is applied on the noisy image to obtain detail sub-bands. Then, the median filter is applied on each sub-band independently using the equations 4, 5 and 6. Then IDWT, MSE and PSNR are calculated as in case one. Figure 4 shows the block diagram for the third case, Median filter before threshold.

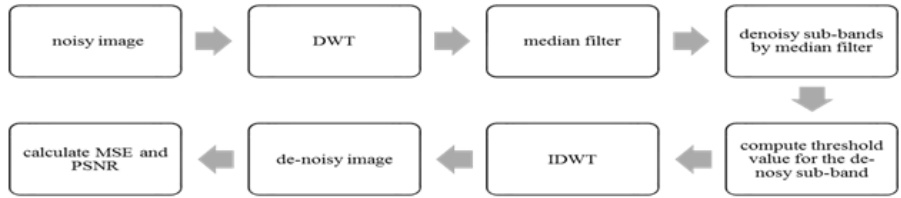

Figure 4: The block diagram for the third case, median filter before threshold.

## 2.4. THE FOURTH CASE: APPLYING MEDIAN FILTER AFTER COMPUTING THE THRESHOLD VALUE

In this case, median filter is applied after computing the threshold value. All steps are as in case one, except the Median filter is applied on the sub-bands after applying threshold value and before the IDWT. Finally, MSE and PSNR values are computed as before. Figure 5 shows the block diagram for the fourth case, Median filter after threshold.

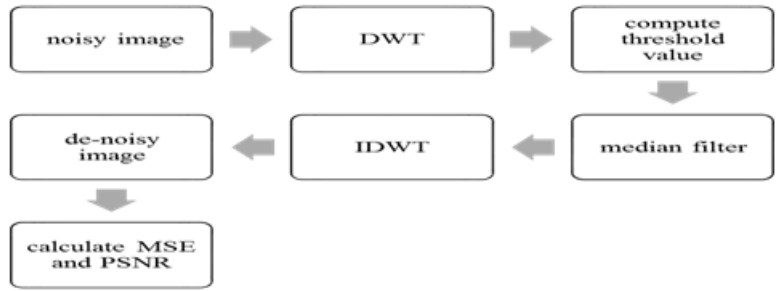

Figure 4: The block diagram for the fourth case, median filter after threshold.

## 3. EXPERIMENTAL RESULTS

The above-mentioned cases of the proposed methods were implemented in MATLAB (R2015a). In this implementation of the algorithm and running the subsequent tests, five standard sample grayscale images sized 256×256 pixels each were used in generating the noisy images. The noise type was Additive Gaussian Noise (AGN) with zero mean (m=0) and the noise intensity values ranging from σ=15, σ=20, and σ=25. The ratio of the additive noise in the noisy image content was unknown by the image de-noising method. For this reason, the Robust Median Estimator was use in order to estimate the amount of noise in the image by the equation 5 [1].

Then, the standard deviation of each sub-band was computed to use it in equation 4 in order to calculate the threshold value for each sub-band by using equation 7 as follows:

$$F'(i, j) = \begin{Bmatrix} F(i, j), & F \geq Th \\ 0, & otherwise \end{Bmatrix}$$

.............. (7)

Where:

$F(i, j)$ : The pixel value of noisy image band.

$F(i, j)$ : The pixel value of de-noised image.

Th : The threshold value.

To calculate the PSNR values, firstly the de-noised image was generated by applying IDWT on the de-noised sub-bands. Such IDWT use was part of all proposed cases above except the second one. The sample original images were Lena, Barbara, Camera, Fruits, and Butterfly. Upon adding AGN in different levels to the original image, and by applying DWT with one stage of analysis on noisy image to obtain high frequency sub-bands HH, HL and LH, in all four cases, the results of the PSNR values are shown in Table 2. Corresponding images are being displayed in Figures 5-9 below.

Table 2: PSNR values for four cases with different noise levels.

| Images | noise ratio | MF | DWT | MF befor thr. | MF after thr. |
|---|---|---|---|---|---|
| Lena | 15 | **26.5469** | 25.1504 | 26.4146 | 26.2893 |
| | 20 | **25.5292** | 23.5935 | 25.1486 | 25.2302 |
| | 25 | **24.6117** | 22.3148 | 24.0477 | 24.1414 |
| Barbara | 15 | 22.3048 | 23.5818 | **24.0892** | 23.2405 |
| | 20 | 21.9372 | 22.3197 | **23.1468** | 22.6969 |
| | 25 | 21.5613 | 21.2449 | **22.2978** | 22.0692 |
| Camera | 15 | 24.6562 | 24.5340 | **25.1232** | 24.4323 |
| | 20 | 24.0361 | 22.9896 | **24.2206** | 23.8242 |
| | 25 | **23.3887** | 21.6859 | 23.2202 | 23.0355 |
| Fruit | 15 | 25.7715 | 24.7774 | **25.9091** | 25.3040 |
| | 20 | **24.9298** | 23.2299 | 24.8211 | 24.4766 |
| | 25 | **24.1101** | 22.1286 | 23.7675 | 23.6186 |
| Buterfly | 15 | 23.4540 | 24.0486 | **24.7068** | 23.7919 |
| | 20 | 22.8565 | 22.5758 | **23.7164** | 23.2238 |
| | 25 | 22.3289 | 21.3814 | **22.8253** | 22.5696 |

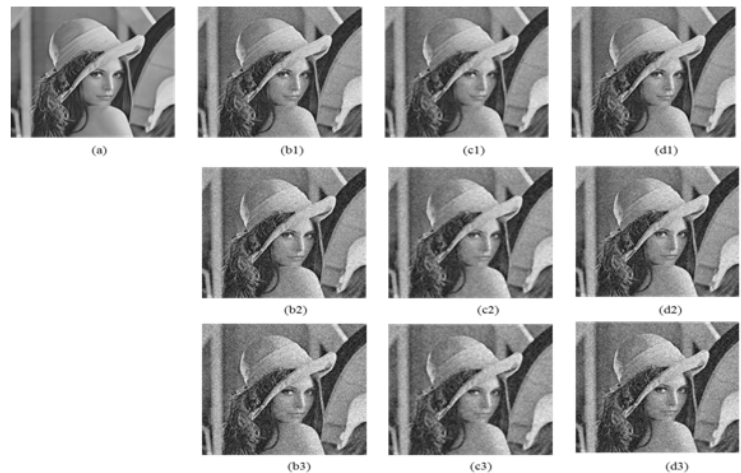

Figure 5: Result of the best PSNR value on de-nosing Lena image by median filter only and applying median filter before threshold, (a) original image  (b1,b2,b3) noisy image corrupted by noise ratio=15,20,25

respectively,(c1,c2,c3) de-noising by median filter only,(d1,d2,d3) de-nosing by applying median filter before threshold

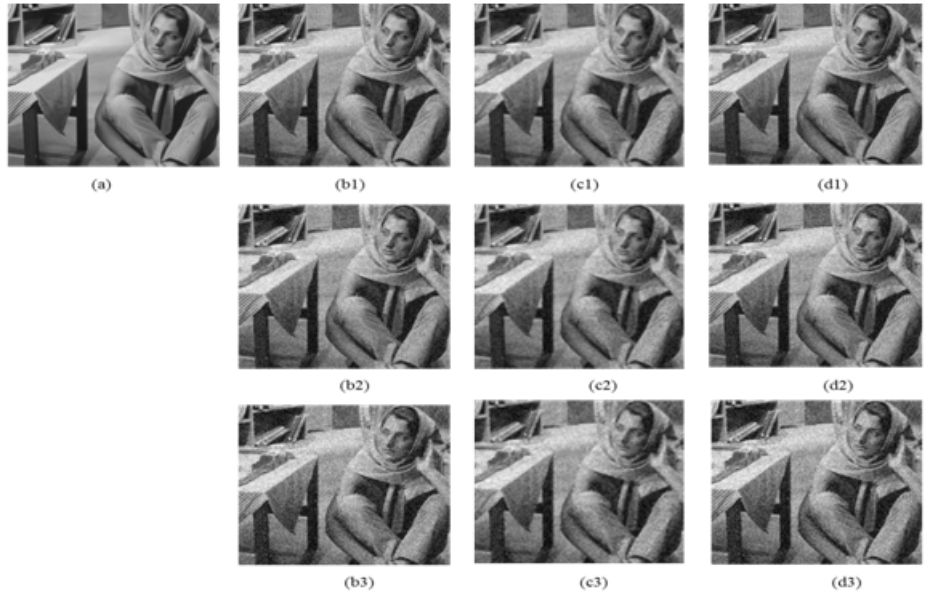

Figure 6: Result of the best PSNR value on de-nosing Barbara image by median filter only and applying median filter before threshold, (a) original image , (b1,b2,b3) noisy image that corrupted by noise ratio=15,20,25 respectively,(c1,c2,c3) de-noising by median filter only,(d1,d2,d3) de-nosing by applying median filter before threshold.

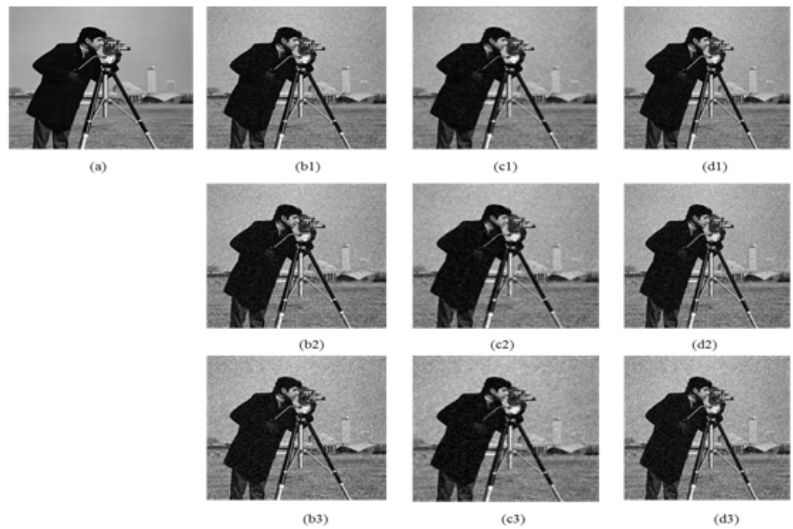

Figure 7: Result of the best PSNR value on de-nosing Camera image by median filter only and applying median filter before threshold, (a) original image , (b1,b2,b3) noisy image that corrupted by noise ratio=15,20,25 respectively,(c1,c2,c3) de-noising by median filter only,(d1,d2,d3) de-nosing by applying median filter before threshold

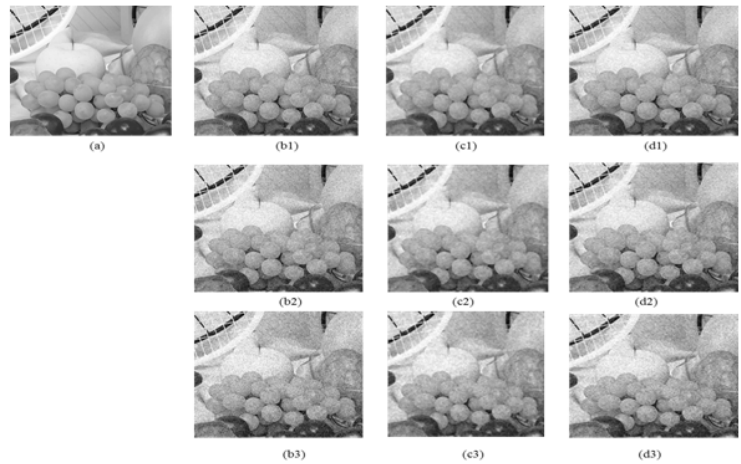

Figure 8: Result of the best PSNR value on de-nosing Fruits image by median filter only and applying median filter before threshold, (a) original image , (b1,b2,b3) noisy image that corrupted by noise ratio=15,20,25 respectively,(c1,c2,c3) de-noising by median filter only,(d1,d2,d3) de-noising by applying median filter before threshold

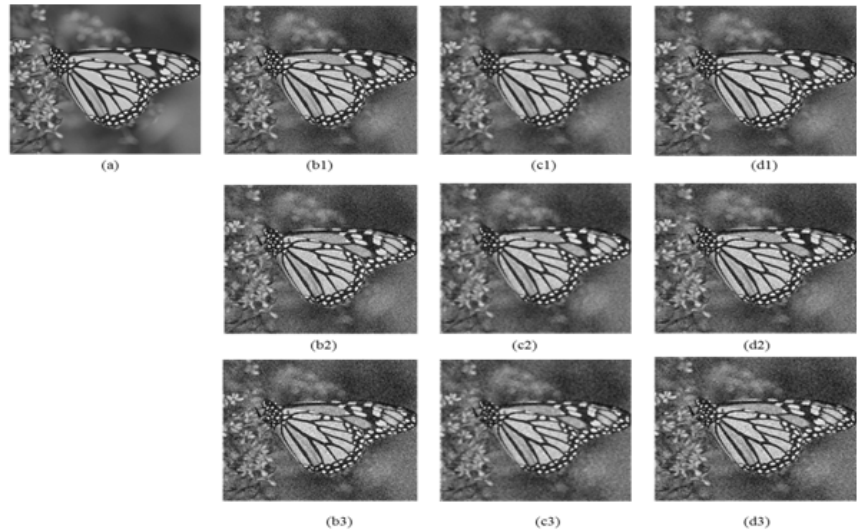

Figure 9: Result of the best PSNR value on de-nosing Butterfly image by median filter only and applying median filter before threshold, (a) original image (b1,b2,b3) noisy image that corrupted by noise ratio=15,20,25 respectively,(c1,c2,c3) de-noising by median filter only,(d1,d2,d3) de-nosing by applying median filter before threshold

## 4. DISCUSSION

The summary of the comparative analysis of PSNR values for five tested images, namely Lena, Barbara, Camera, Fruits and Butterfly, at different level noises, are being displayed in table 2 and the images themselves in figures 5,6,7,8, and 9. The efficiency of the proposed de-noising methods was dependent on the nature of the image; for example, applying only median filter has a good effect on Lena, Camera and Fruits images. Their images reveal a smooth texture. However, the proposed methods of applying median filter before and after threshold on Barbara and Butterfly produces coarse texture images. Therefore, median filter before and after threshold have a better performance and efficiency than applying DWT only, which showed a poor effect compared with other methods. This finding is in agreement with Kumar and Agarwal [10]. We concluded that hybrid filter provides clear images and visually better quality. The mix between using DWT and the median filter have shown significant results and they were able to recover much more detail of the original image, providing a successful way of image de-noising. This finding was consistent with the previous study by Shukla et al [16].

## 5. CONCLUSIONS AND FUTURE WORK

In this research, a new method of image de-noising is proposed. The proposed method employs median filter and adaptive wavelet threshold, using different types of wavelet transform filters. It achieves better PSNR values compared to that of applying DWT or median filter only. The quality of images produced using the proposed de-noising method on the images corrupted with Gaussian noise was compared in terms of PSNR against classic methods for DWT and the median filter each alone. According to experimental results, the proposed method presents best values of PSNR for the de-noised images. The robust median estimator was used to estimate the noise ratio in the noisy images.

In the future, the proposed method can be modified by using more accurate estimation function and other kinds of threshold methods in trying to improve its ability at image de-noising process.

## REFERENCES

[1] Liu, Y. (2015). Image De-noising Method based on Threshold, Wavelet Transform and Genetic Algorithm. International Journal of Signal Processing, Image Processing and Pattern Recognition, 8(2), 29-40.

[2] Kaur, G., & Kaur, R. (2012). Image De-noising Using Wavelet Transform and Various Filters. International Journal of Research in Computer Science, 2(2), 15.

[3] Charde, M. P. (2013). A Review on Image De-noising Using Wavelet Transform and Median Filter over AWGN Channel. Vol. 1, 44-47

[4] Burrus C.S., Gopinath R.A., Guo H. (1998). Introduction to Wavelets and Wavelet Transforms: A Primer. Prentice Hall, USA, 281p.

[5] Toufik, B., & M Okhtar, N. (2012). The Wavelet Transform for Image Processing Applications. INTECH Open Access Publisher. 396-422

[6] Ergen, B. (2012). Signal and image de-noising using wavelet transform. INTECH Open Access Publisher. pp 496-514

[7] Om, H., & Biswas, M. (2012). An improved image de-noising method based on wavelet thresholding. Journal of Signal and Information Processing, Vol. 3 No. 1, 2012, pp. 109-116

[8] Chang, S. G., Yu, B., & Vetterli, M. (2000). Adaptive wavelet thresholding for image de-noising and compression. IEEE transactions on image processing, 9(9), 1532-1546.

[9] Hedaoo, P., & Godbole, S. S. (2011). Wavelet thresholding approach for image de-noising. International Journal of Network Security & Its Applications (IJNSA), 3(4), 16-21.

[10] Kumar, P., & Agarwal, S. K. (2015). A Color Image Denoising By Hybrid Filter for Mixed Noise. Journal of Current Engineering and Technology, 5(3). 1565-1572

[11] Saxena, C., & Kourav, D. (2014). Noises and Image De-noising Techniques: A Brief Survey. International Journal of Emerging Technology and Advanced Engineering, 4(3), 878-885.

[12] Gupta, B., & Negi, S. S. (2013). Image Denoising with Linear and Non-Linear Filters: A Review. International Journal of Computer Science, 10(2), 149-144.

[13] Qian, W. (2015). Research on Image De-noising with an Improved Wavelet Threshold Algorithm. International Journal of Signal Processing, Image Processing and Pattern Recognition, 8(9), 257-266.

[14] Tai-sheng, Z. (2015). Research on Image Denoising with Wavelet Transform and Finite Element Method. International Journal of Signal Processing, Image Processing and Pattern Recognition, 8(10), 363-374.

[15] Ismael, S. H., Mustafa, F. M., & Okümüs, I. T. (2016, March). A New Approach of Image Denoising Based on Discrete Wavelet Transform. In Computer Applications & Research (WSCAR), 2016 World Symposium on (pp. 36-40). IEEE.

[16] Shukla, H. S., Kumar, N., & Tripathi, R. P. (2014). Median Filter based Wavelet Transform for Multilevel Noise. International Journal of Computer Applications, 107(14).

## AUTHORS

**Afrah Ramadhan** master's student in Department of Electrical-Electronics and Computer Engineering at Aksaray University Aksaray, Turkey

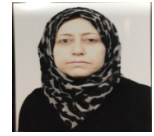

**Dr. Firas Mahmood Al-Fiky**, received his BSc, MSc and PhD degrees from University of Mosul-Iraq, College of Engineering, Electric dept. at 1997, 2000 and 2007 respectively. He has a PhD in Computer Engineering, he is lecturer since 2007. He started teaching and supervising post graduate courses since 2009 and until now. He is a head of petrochemical engineer in Duhok Polytechnic University (DPU).

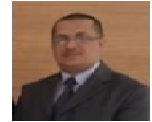

**Atilla Elçi** is full professor and the chairman of Department of Electrical-Electronics and Computer Engineering at Aksaray University since August 2012. He has served in Computer Engineering departments in various universities including METU, Turkey and EMU, TRNC, since 1976. His professional practice includes the International Telecommunication Union (ITU), Switzerland, as chief technical advisor for field projects on computerization of telecommunication administrations of member countries (1985-97) and Information Technology & Telecommunications Pvt Ltd as founder/managing director, Turkey (1997-2003). He has organized IEEE COMPSAC & ESAS since 2006, SIN Conferences since 2007; IJRCS Symposiums 2008-9, ICPCA_SWS 2012 and, LightSec 2016. He has published over a hundred journal and conference papers and book chapters; co-authored a book titled The Composition of OWL-S based Atomic Processes (LAP Lambert, 2011); edited the Semantic Agent Systems (Springer, 2011), Theory and Practice of Cryptography Solutions for Secure Information Systems (IGI Global, 2013), The Handbook of Applied Learning Theory and Design in Modern Education (IGI Global, 2016), Metacognition and Successful Learning Strategies in Higher Education (IGI Global, 2017) and the proceedings of SIN Conferences 2007-16 (ACM), ESAS 2006-16 (IEEE CS). He is an associate editor of Expert Systems: The Journal of Knowledge Engineering and editorial board member of MTAP, JSCI, IJAS, IJISS, and guest editor for several other journals. He has delivered several keynote/invited talks and served as reviewer for numerous conferences.

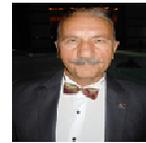